\documentclass[12pt]{article}
\usepackage{graphicx,epsf}
 \advance\oddsidemargin by -1in
 \advance\evensidemargin
 by -1in
 \marginparsep
 0.4cm
 \advance\topmargin by
 -0.0in
 \makeindex

 \newcommand\beq{\begin{equation}}
 \newcommand\eeq{\end{equation}}
 \newcommand\beqn{\begin{eqnarray}}
 \newcommand\eeqn{\end{eqnarray}}
\newcommand{\la}{\langle}
 \newcommand{\ra}{\rangle}

 \def\mb{\,\mbox{mb}}
 \def\fm{\,\mbox{fm}}
 \def\GeV{\,\mbox{GeV}}
 \def\GeV/c{\,\mbox{GeV/c}}
 \def\MeV{\,\mbox{MeV}}
 \def\MeV/c{\,\mbox{MeV/c}}
 \def\lsim{\mathrel{\rlap{\lower4pt\hbox{\hskip1pt$\sim$}}
    \raise1pt\hbox{$<$}}}
 \def\gsim{\mathrel{\rlap{\lower4pt\hbox{\hskip1pt$\sim$}}
    \raise1pt\hbox{$>$}}}

\def\mb{\,\mbox{mb}}
\def\fm{\,\mbox{fm}}
\def\GeV{\,\mbox{GeV}}

\begin{document}

  \title
  {\bf Time Evolution of Hadronization\\ 
and Grey Tracks in DIS off Nuclei}
  \author{\large C. Ciofi degli Atti$^a$ and B.Z. 
Kopeliovich$^{b,c}$}
 \maketitle

\begin{center}
  {$^a$ Department of Physics, University of Perugia, and
\\ INFN, Sezione di Perugia, via A. Pascoli, Perugia, I-06100, Italy\\
$^b$Max-Planck-Institut f\"ur Kernphysik, Postfach 103980, 69029
Heidelberg, Germany\\
$^{c}$Joint Institute for Nuclear Research, Dubna, Russia
}
\end{center}

   \begin{abstract}

The analysis of the grey tracks produced in Deep Inelastic Scattering
(DIS) off nuclei provides important information on the space-time
development of hadronization in nuclear medium. This method is
complementary to the measurement of nuclear attenuation of leading
inclusive hadrons. While the latter is focused on the hadronization
dynamics for the quite rare process of leading hadrons production, the
former covers the main bulk of inelastic events, and its $Q^2$ dependence
is a very sensitive tool to discriminate between different models of
hadronization. Employing the model of perturbative hadronization
developed earlier, we calculate the $Q^2$ and $x_{Bj}$ dependences of the
number of collisions and relate it to the mean number of grey tracks,
using an empirical relation obtained from the analysis of data from the
Fermilab E665 experiment on DIS of muons off the $Xe$ nucleus.  We found
the number of grey tracks to rise steeply with $Q^2$ in good agreement
with the experimental data.
   
   \end{abstract}


  \newpage

 Nuclear targets serve as a natural and unique analyzer of the space-time
development of strong interactions at high energies \cite{knp,knph}. Due
to Lorentz time dilation, the projectile partons may keep their coherence
for some time, but once they become incoherent, the cross section of the
final state interaction (FSI) increases and the nucleus may act as a
filter of hadronization mechanisms, provided quantities which are
sensitive to the nuclear modifications of FSI are measured.

One way for experimental testing of various theoretical models which have
been proposed to investigate the space-time evolution of hadronization,
is the measurement of the nuclear modification factor in inclusive
production of leading hadrons \cite{emc,slac,hermes}. Recent experimental
data from the HERMES experiment at HERA \cite{hermes} have confirmed the
predictions made in \cite{knp}, based on the perturbative gluon radiation
approach.  Attenuation of the leading hadron is partially ascribed to the
FSI of the pre-hadron (a colorless $\bar q q$ dipole prior formation of
the hadronic wave function) produced in the color neutralization of the
radiating quark which was created in the hard ${\gamma^*}$-nucleon
interaction. However, other approaches for in-medium hadronization have
recently been employed, which are also able to provide a reasonable
explanation of the HERMES data. We mean, in particular, the approach of
Ref.~\cite{ww}, in which nuclear attenuation is explained in terms of the
induced energy loss scenario, in which the hit parton loses energy by
hard multiple scattering with other nucleon's partons, and the approaches
of Ref. \cite{amp,giessen,ako}, based upon the color string model. Thus
it appears that other, more selective phenomena which could discriminate
various models of hadronization are required. In this respect, the
investigation of the $Q^2$ dependence of nuclear effects would probably
be one of the best candidates \footnote{In this note $Q^2=-q^2={\bf q}^2
- \nu^2$ denotes the squared four-momentum transfer}.  As a matter of
fact, recent data from HERMES are in a good agreement with calculations
performed in \cite{knph}, but in strict contradiction with the $Q^2$
dependence predicted in \cite{ww}. Unfortunately the HERMES data exhibit
only a mild $Q^2$ dependence and therefore could not be very selective
for other hadronization models.

 Note that production of leading hadrons with large fraction $z_h$ of the
initial parton momentum is a rare, nearly exclusive process, which has
quite a specific time development. In the main bulk of events, the jet
energy is shared by many hadrons and therefore in this case it is a
difficult task to find observables which are sensitive to the time
development of hadronization. In Ref.~\cite{ck} it has been proposed as a
possible candidate the DIS exclusive process $A(e,e'B)X$, where $B$ is a
detected recoil nucleus with a less number of nucleons than the target one
\cite{cks}, and $X$, as usual, denotes the unobserved jets of hadrons
resulting from hadronization. It has indeed been shown \cite{ck} that the
survival probability of the recoil nucleus $B$ in such a process is rather
sensitive to the time evolution of hadronization and, particularly, to its
$Q^2$ dependence. This quantity, which is quite difficult to access
experimentally in case of heavy nuclei, could be investigate using
few-body nuclei. Indeed, first experimental data were recently obtained at
Jefferson Lab \cite{khun} for a deuteron target, i.e. by measuring the
recoiling protons in the DIS process $D(e,e'p)X$ in a wide range of
kinematics. Preliminary results of calculations \cite{ckk} show a good
agreement with the predictions of the hadronization model of
Ref.~\cite{ck}.

It should however be pointed out that the dominant channels of DIS are
associated with events where the recoil nucleus $B$ does not survive, but
breaks apart to fragments, whose distribution in $Q^2$ and other
kinematic variables could also shed light on the time evolution of the
jet in the nuclear matter. A typical process of this type is the
production of the so called "grey tracks" (GT) in DIS semi-inclusive
processes. GT are those hadrons, predominantly protons, whose momenta are
in a few hundred $\MeV/c$ range. They have been observed in various DIS
processes induced by different projectiles; in this note we will consider
the GT production in the Fermilab $E665$ experiment \cite{e665} on
$\mu-Xe$ and $\mu-D$ processes at $490\GeV$ beam energy; in this
experiment the GT have been associated to protons with momentum in the
interval $200-600\MeV/c$.

The investigation of the $Q^2$-dependence of the GT production may bring
forth precious information about multiple FSI of the jet originated from
the DIS in the nuclear medium. The mean number of GT, to be denoted
henceforth as $\la n_{gr}\ra$, correlates with the intensity of the FSI:
this is why the production of GT has been traditionally used as a probe
for the dynamics of the interaction and the centrality of collisions. The
latter is usually characterized by the mean number of collisions,
$\la\nu_c\ra$, i.e.  by the number of bound nucleons which took part in
the interaction \footnote{Note that this number is different from the
expansion parameter, $\sigma_{in}T_A(b)$ in the Glauber formula which is
also frequently called number of collisions (see the discussion in
\cite{mine})}.

The dependence of $\la n_{gr}\ra$ upon $Q^2$, $\nu$ and $x_{Bj}$ has been
measured in the E665 experiment and it is the aim of this note to present
a theoretical interpretation of such a dependence resulting from a
specific model of hadronization. To this end we will proceed in the
following way: our hadronization model allows one to calculate the average
number of collisions $\la\nu_c\ra$ which is related to $\la n_{gr}\ra$.  
The relation between the observed $\la n_{gr}\ra$ and models for
$\la\nu_c\ra$ can be obtained with various Monte-Carlo generators for
hadron cascading in nuclei. The results of these calculations however do
not seem to be very accurate and a more reliable and model independent
approach has been adopted in \cite{e665}, where the following relation has
been found,
 \beq
\la\nu_c\ra = (2.08\pm0.13) +(3.72\pm0.14)\,\la n_{gr}\ra
\ ,
\label{10}
 \eeq 
 basing on the measured average total hadronic net charge and its relation 
to $\la\nu_c\ra$.

We have used the same relation Eq.~(\ref{10}) to obtain $\la n_{gr}\ra$,
after calculating the average number of collisions $\la\nu_c\ra$ occurring
during the evolution of a jet produced in DIS and propagating through
nuclear matter. For minimum bias DIS events we obtain,
 \beq
\la\nu_c\ra = \int d^2b\,
\int\limits_{-\infty}^\infty dz\,\rho_A({\bf b},z)
\int\limits_z^\infty dz'\,\rho_A({\bf b},z')\,
\sigma_{eff}(z-z')\,+\,1\ .
\label{20}
 \eeq
 This equation results from the following description of the DIS process:
the hard interaction of the lepton takes place on a bound nucleon at
impact parameter ${\bf b}$ and longitudinal coordinate $z$; then the
knocked out quark hadronizes and initiates a jet which propagates through
the nucleus interacting with other bound nucleons with the effective cross
section $\sigma_{eff}(z-z')$ depending on the distance (or time, provided
that the quark propagates with the speed of light). The first term in
Eq.~(\ref{20}) gives the amount of new nucleons involved in the process,
i.e. the number of collisions, whereas the second term represents the
contribution from the recoil nucleon formed in the initial, hard collision
between ${\gamma^*}$ and one of the bound nucleons.

Evaluation of Eq.~(\ref{20})  requires an explicit form for the
time-dependent effective cross section $\sigma_{eff}$, which can only be
obtained within a model for hadronization. To this end, we employ the
model suggested in \cite{ck}: it combines the soft part of the
hadronization dynamics, described in terms of the string model, with the
hard part, described within perturbative QCD. There are many experimental
evidences showing that gluons are located within small clouds around the
valence quarks \cite{kst2,kp} .  Therefore, if $Q$ is less than the
mass scale $\lambda=0.65\,GeV$ controlling the transverse quark-gluon
separation, they can hardly be shaken off;  in this case the string model
is a proper description of hadronization. If, however, $Q>\lambda$,
perturbative gluon radiation should be taken into account. Employing
$\lambda$ as the bottom limit for the integration over the gluon
transverse momenta (see below), double counting is excluded when the
string and pQCD contributions are added to describe hadron production.
This model of hadronization is close to the one used in \cite{knp,knph},
where the string contribution was mocked up by the soft part of gluon
spectrum, and the predictions of the two versions of the model are 
rather similar.

It was found in \cite{ck} that the multiplicity of the produced 
hadrons, or better  to say  the pre-hadrons  which are colorless dipoles,
rises with time as
 \beq
\la n_h(t)\ra = n_M(t) + n_G(t)\ ,
\label{40}
 \eeq
where  $n_M$  is the amount of the pre-hadrons produced due to decays of
the string,  and $n_G(t)$ the one produced by gluon radiation.

The string contribution to $\la n_h(t)\ra$ has been found in \cite{ck}
employing the standard dynamics of string decay \cite{cnn,k-pl}. The
pre-hadron multiplicity as function of time, $n_{M}(t)$, was found in the
following form \cite{ck},
 \beq
n_{M}(t)=
\frac{{\rm ln}(1+t/\Delta t)}
{{\rm ln}2}. 
\label{50}
 \eeq
 Here the time scale $\Delta t$ is related to the probability $w$ of a
$\bar qq$ pair creation in the color field of the string, per unit of time
and per unit of string length. This parameter evaluated either in the
Schwinger model or from the phenomenology turns out to be $w\approx
2\fm^{-2}$. Correspondingly, $\Delta t=\sqrt{2/w}=1\fm$.

Note that the logarithmic dependence on time in Eq.~(\ref{50}) is rather
obvious. It is related to the fact that hadrons produced via string decays
build a plateau in rapidity. Since the momenta acquired by pre-hadrons are
proportional to the time taken by the string decay, the number of decays
rises linear in the $\ln(t)$ scale.

Concerning the gluon contribution to the hadron multiplicity
Eq.~(\ref{40}), we employ the large $N_c$ approximation and replace each
radiated gluon by a color octet $\bar qq$ pair, and then combine the
quarks and antiquarks into colorless dipoles (pre-hadrons). This is the
origin of the second term $n_G(t)$ in (\ref{40}).

It is well known that radiation does not happen instantaneously, but takes
time called coherence time, $t_c=2\omega/k_T^2$, where $\omega$ and $k_T$
are the energy and transverse momentum of the radiated gluon. At shorter
times the quark-gluon system is still in coherence with the initial quark.
Integrating the perturbative gluon radiation spectrum over $\omega$ and
$k_T$ with a weight factor $\Theta(t-t_c)$ one gets the amount of gluons
which lost coherence, i.e. were radiated over time interval $t$ after the
DIS interaction.  The time dependence of the gluon radiation was found in
\cite{ck} to be controlled by the parameter $t_0=(m_N\,x_{Bj})^{-1} =
0.2\fm/x_{Bj}$, and the number of perturbatively radiated gluons was
evaluated in the following form
 \beq
n_G(t) = \frac{16}{27}\,\left\{
{\rm ln}\left(\frac{Q}{\lambda}\right)\,+\,
{\rm ln}\left(\frac{t\,\Lambda_{QCD}}{2}
\right)\,{\rm ln}\left[\frac{{\rm ln}(Q/\Lambda_{QCD})}
{{\rm ln}(\lambda/\Lambda_{QCD}}\right]\right\}\ ,
\label{60}
 \eeq
 for $t < t_0$. At longer times,
$t > t_0$, the $t$-dependence starts leveling off,
 \beqn
n_G(t) &=& \frac{16}{27}\,\left\{
{\rm ln}\left(\frac{Q}{\lambda}
\,\frac{t_0}{t}\right)\,+\,
{\rm ln}\left(\frac{t\,\Lambda_{QCD}}{2}
\right)\,{\rm ln}\left[\frac{{\rm ln}(Q/\Lambda_{QCD}
\sqrt{t_0/t})}
{{\rm ln}(\lambda/\Lambda_{QCD})}\right]
\right.\nonumber\\ &+& \left.
{\rm ln}\left(\frac{Q^2\,t_0}{2\,\Lambda_{QCD}}
\right)\,{\rm ln}\left[\frac{{\rm ln}(Q/\Lambda_{QCD})}
 {{\rm ln}(Q/\Lambda_{QCD}\,\sqrt{t_0/t})}\right]\right\}
\ ,
\label{70}
 \eeqn
 and saturates at $t > t_0\,Q^2/\lambda^2 = 2\nu/\lambda^2$, which is a
very long time interval, substantially exceeding the nuclear size for the
energies $\nu=50-400\GeV$ covered in the E665 experiment.

Having obtained the explicit forms of the string and perturbative
radiation contributions in Eq. (\ref{40}), let us now discuss the
interaction cross section of the produced colorless dipoles. These, as we
mentioned, should be treated as pre-hadrons, until they form hadronic wave
functions and get definite masses. As for their interaction cross section,
what matters is their transverse size, rather than their mass. Since the
decay of a string is a soft process, it is natural to assume that the
decay products have a size of the order of the hadronic size and therefore
interact with about the same cross section.

Less obvious is the situation with pre-hadrons produced from perturbative
gluons. Since the mean transverse momenta of the gluons follow the photon
virtuality $Q^2$, the initial size of the produced pre-hadrons should be
small, of the order of $1/Q$. Then one may expect the color transparency
effect to be at work \cite{zkl,bm}.  Indeed, this happens for leading
hadrons, and the evolution of the pre-hadron size has a strong impact on
nuclear transparency \cite{knp,knph}. In the present case, however, we
deal with the slowest part of the hadron spectrum which is independent of
the photon energy $\nu$, provided that it is sufficiently high.  Only
those pre-hadrons are produced within the nuclear volume whose energy
does not exceed $E_h \lsim R_A\,(Q^2-\lambda^2)/2\ln(Q^2/\lambda^2)$. The
size of such a pre-hadron evolves so fast that the largest part of the
effect of color transparency is washed out. Even for the most energetic
exclusively produced hadrons in the E665 experiment, the effect of color
transparency is rather moderate \cite{e665-ct,knnz}; it is much weaker
and hardly detectable at lower energies \cite{knst}. Therefore, we
neglect these corrections and assume that pre-hadrons interact with the
hadronic cross sections.

Eventually, we are in the position to evaluate the time dependent
effective absorption cross section in (\ref{20}), to be used in the
description of GT production; by treating all the ${\bar qq}$ colorless
dipoles as mesons (M), we obtain
 \beq
\sigma_{eff}(t)=
\sigma^{MN}_{in}\Bigl[n_M(t) +
n_G(t)\Bigr]\ ,
\label{80}
 \eeq

 Note that this expression is different from the one used in \cite{ck}.
Firstly, we excluded the term $\sigma^{NN}_{tot}$ corresponding to
interaction of the nucleon originated from the first DIS event, since we
treat this nucleon as a participant giving rise to the last term in
(\ref{20}).  Secondly, we replaced the total meson-nucleon (MN) cross
section by the inelastic one. This is because most of recoiling protons
in elastic and diffractive scattering have too small momenta to
contribute to the production of GT. Therefore, we use $\sigma^{MN}_{in} =
\sigma^{\pi N}_{tot} - \sigma^{\pi N}_{el} - \sigma^{\pi N}_{sd} =
17.7\mb$.

Before illustrating the calculation of the $Q^2$ dependence of the mean
number of GT $n_{gr}(Q^2)$, let us briefly discuss the expected
dependence of $n_{gr}$ upon the energy transfer $\nu$. Both the string
model and perturbative radiation lead to the mean multiplicity which
rises logarithmically with energy.  Naively, one may conclude that the
amount of GT follows the multiplicity and should rise with energy as
well.  However, as already discussed, it takes time to produce hadrons.
The observed multiplicity does not emerge in an instantaneous
explosion-like particle production, but is a result of time development
of the hadronization process. The full time of jet production is
proportional to its energy and exceeds the nuclear size at the E665
energies.  However, only those gluons which are radiated inside the
nucleus can contribute to the production of GT. With the condition
$t_c<R_A$ one should integrate the radiation spectrum from
$\omega_{min}=k_T$ up to $\omega_{max}=k_T^2 R_A$. The result is
independent of the jet energy. Indeed, this is confirmed by the results
of the E665 experiment (see Figs.~2, 17 in \cite{e665}).

At the same time, one should expect a $Q^2$ dependence of the mean number
of GT. Indeed, the stronger kick gets the quark, the more gluons it shakes
off. The mean transverse momentum of radiated gluons follows $Q^2$ which
plays the role of the upper cut-off for the integration $d^2k_T/k_T^2$.
Correspondingly, the multiplicity of the radiated gluons rises
logarithmically with $Q^2$.

\begin{figure}[tbh] 
\includegraphics{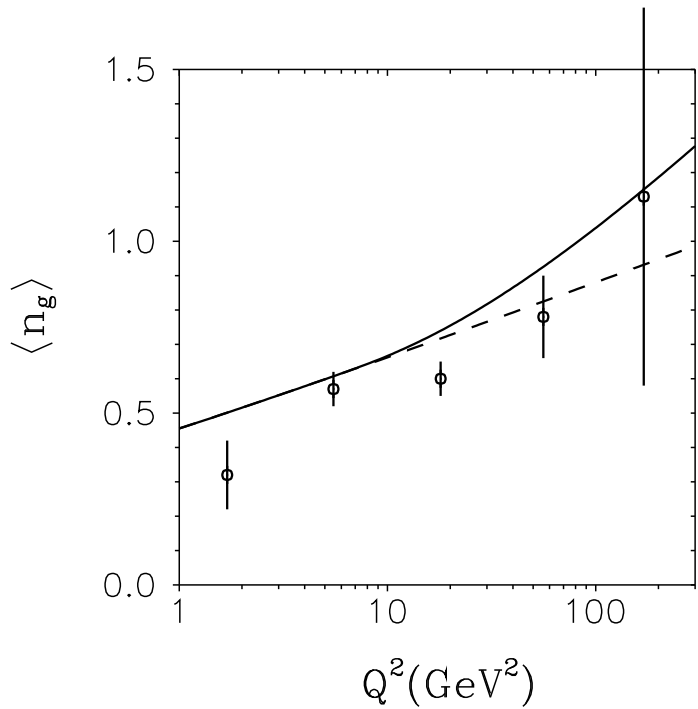}
\begin{center}                  
\vspace{9.5cm}
\parbox{13cm}
{\caption[Delta]
 {The mean number of grey tracks $\la n_{gr}\ra$ produced in the $\mu Xe$
DIS as function of $Q^2$. The theoretical predictions based upon
Eqs.~(\ref{10}), (\ref{20}) and (\ref{80}), are compared with the
experimental data from Ref. \cite{e665}, which correspond to non-shadowing
($x_{Bj}>0.02$) region. The dashed curve corresponds to fixed $x_{Bj}=\la
x_{Bj}\ra = 0.068$. The solid curve includes the correlation between $Q^2$
and $x_{Bj}$ (see text).}
 \label{gt}}
\end{center}
 \end{figure}

Let us now quantitatively analyze the behavior of $\la n_{gr}(Q^2)\ra$.  
To this end we have calculated the mean number of collisions by using
Eqs.~(\ref{20}) and (\ref{80}). Then, applying (\ref{10}), we have
obtained the number of GT, $\la n_{gr}(Q^2)\ra$. We fixed $x_{Bj}$ at the
mean value $\la x_{Bj}\ra=0.068$ \cite{e665}.  The result shown by dashed
curve is compared with the experimental data from the E665 experiment
\cite{e665} in Fig.~\ref{gt}.
 
If the time interval $t_0$ is of the order or smaller than the nuclear
size, the chosen value of $x_{Bj}$ affects hadronization and is
important. The E665 data are subject to quite a strong correlation
between $Q^2$ and $x_{Bj}$, which may change the results depicted in
Fig.~\ref{gt}. We parametrize this correlation as $Q^2=A+B\,x_{Bj}$,
where the parameters $A=2.2\GeV^2$ and $B=178\GeV^2$ are found using the
values of $\la Q^2\ra$ and $\la x_{Bj}\ra$ for two regions called in
\cite{e665} "shadowing" and "non-shadowing".  At very small $x_{Bj}<0.01$
we introduce an additional $x_{Bj}$ dependence of parameter $A$ which is,
however, unimportant for the kinematic region under discussion. Our
results corrected for the $Q^2-x_{Bj}$ correlation are depicted by the
solid curve in Fig.~\ref{gt}. In spite of the difference between the two
curves, both agree with the data.

\begin{figure}[tbh] 
\includegraphics{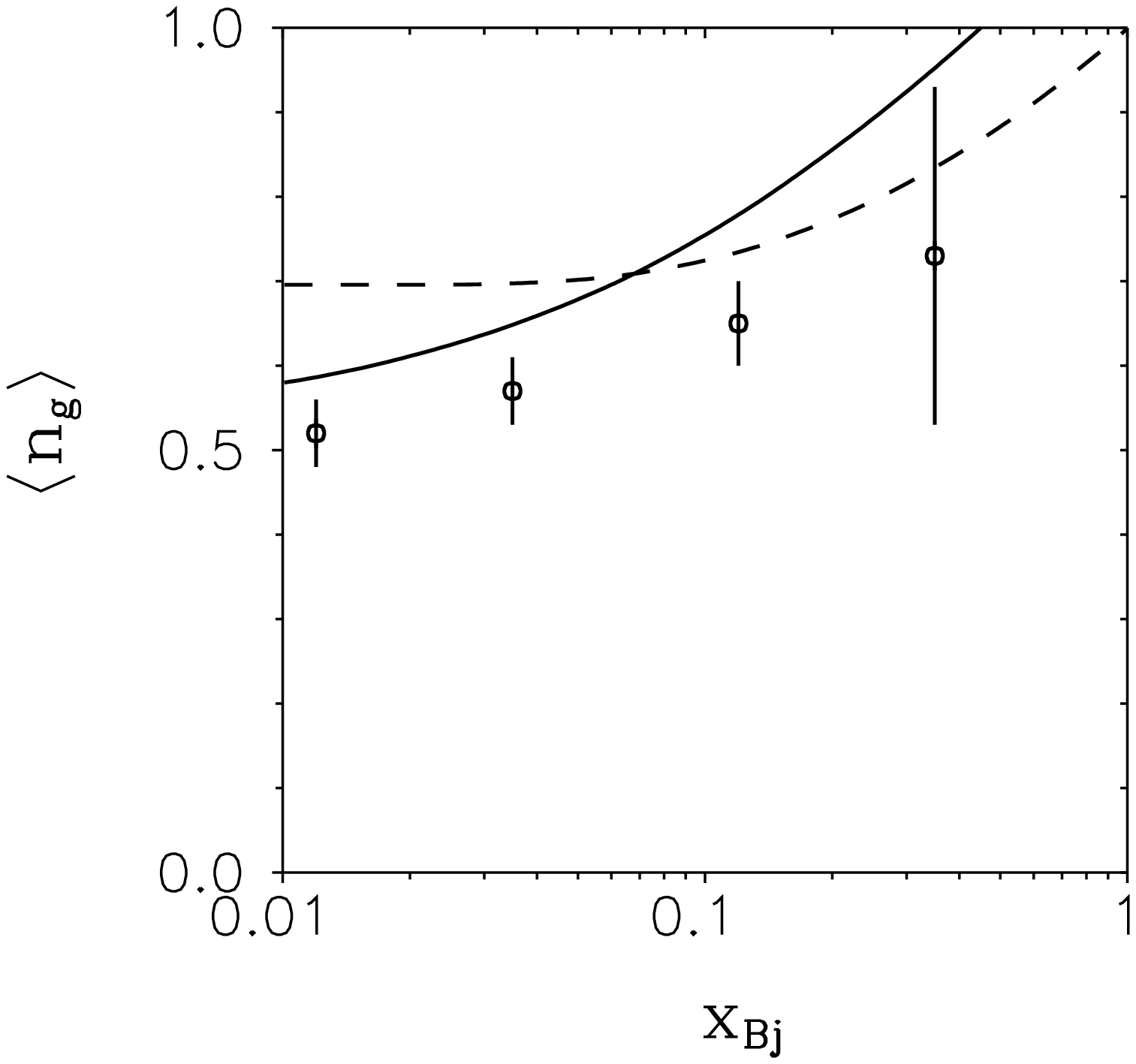}
\begin{center}                  
\vspace{9.5cm}
\parbox{13cm}
{\caption[Delta]
 {The mean number of grey tracks $\la n_{gr}\ra$ produced in the $\mu Xe$
DIS as function of $x_{Bj}$.  The experimental data are from Ref.
\cite{e665}. The dashed and solid curves show the results of calculations
with fixed $Q^2=14.3\GeV^2$ and corrected for the $Q^2-x_{Bj}$ correlation
respectively.}
 \label{xbj}}
\end{center}
 \end{figure}

In order to see the dependence of $\la n_{gr}\ra$ on $x_{Bj}$ explicitly
we calculated it either fixing $Q^2$ at the mean value $\la
Q^2\ra=14.3\GeV^2$, or applying the $Q^2-x_{Bj}$ correlation introduced
above. The results are depicted in Fig.~\ref{xbj} by the dashed and solid
curves, respectively.
 Although both curves slightly overestimate the data, the solid one 
better reproduces the measured slope. 

In closing, the following comments are in order:
 \begin{itemize}
 \item
 In our calculations we employed the same hadronization model as in
\cite{knp,knph,ck} with no readjustment of the parameters, and the
empirical relation between the number of grey tracks and the number of
collisions given by Eq.(\ref{10}). The model successfully passed the new
test;
 \item
 the results of calculations may be subject to further corrections. In
particular, we neglected resonance decays, which is justified by their
rather high energies. On the other hand, the formation time of hadronic
wave function is much shorter than the resonance life time and is rather
short for pre-hadrons produced within the nucleus. This is the reason why
we neglected the effects of color transparency which are important for
leading hadrons \cite{knp,knph}. This may explain why our results
exhibited in Fig.~\ref{gt} somewhat overestimate the data;
 \item
 the $Q^2$ dependence of the mean number of grey tracks serves as a
sensitive tool for testing theoretical models of hadronization. It can be
seen that the amount of grey tracks doubles within the range of $Q^2$
covered by the kinematics of the E665 experiment, and our calculations
well reproduce such a steep variation of $\la n_{gr}(Q^2)\ra$;
 \item 
 whereas nuclear attenuation of leading hadrons carries information of
space-time development of hadronization in rather rare events, grey tracks
provide precious information about hadronization dynamics for the main
bulk of DIS events; 
 \item
 We found a rather flat dependence of $\la n_{gr}\ra$ on $x_Bj$ at fixed
$Q^2$. However, the $Q^2 - x_{Bj}$ correlation existing in the E665
experiment leads to a growth of the number of GT with $x_{Bj}$. Our
calculations corrected for this effect well describe the measured
$x_{Bj}$ dependence of GT;
 \item
 in our calculations we assumed that the value of Bjorken scaling
variable $x_{Bj}$ is sufficiently large to neglect the sea, and this is
the reason why we compared our results with the E665 data at large values
of $x_{Bj}$. At small values of $x_{Bj}$ one should take into account the
production of two jets with different momenta, with only one of them
producing a grey track corresponding to the target nucleon on which the
DIS process occurred. On the other hand, the E665 experiment detected no
substantial change in behavior of different observables in the shadowed
compared to the nonshadowed regions. This was explained by a large
contribution from diffraction \cite{e665}.
 \end{itemize}

\bigskip
 {\bf Acknowledgments:} One of us (C.C.d.A) is grateful for hospitality
and support to the Max-Planck-Institut f\"ur Kernphysik, Heidelberg, where
this paper was initiated, whereas the other (B.Z.K.) is grateful to the
RIKEN BNL Research Center for hospitality during the workshop "Theory
Summer Program on RHIC Physics" when the paper was completed.

\end{document}